# The orientation of historic churches in Las Palmas de Gran Canaria and its surroundings: preliminary results

*A Gangui[1], J Ricra[2], J Pallarés[3], J A Belmonte[4]*

*1. Instituto de Astronomía y Física del Espacio (CONICET-UBA), Argentina*
*2. Facultad de Ciencias, Universidad Nacional de Ingeniería, Perú*
*3. Independent researcher, Gran Canaria, Spain*
*4. Instituto de Astrofísica de Canarias, Tenerife, Spain*

## Abstract

We present preliminary results of the analysis of the precise spatial orientation of colonial Christian churches located in the city of Las Palmas, capital of the island of Gran Canaria (Spain), and its surroundings. Our sample consists of 27 churches, 15 of which belong to the city proper while the rest are situated within roughly 12 km away inland. Most of these churches were constructed between the first decades after the Castilian conquest of the island, and the end of the 19th century. In our analysis we examine whether these orientations respond to calendrical coincidences, to the main features of the landscape, or whether the standard tradition was followed regarding the orientation of the churches' apses eastwards. We find that no church in the sample is oriented towards points of the horizon on which the Sun rises on the day of the patronal feasts. Furthermore, we find no correlation between the global orientation pattern and the orography of the island which, near the local prominent Guiniguada ravine, could have been an important factor at the time of the construction of the churches. What we do find is that most of the orientations fit within the solar range, mainly to the east but also with a few historic churches oriented towards the west, while less than 20% have astronomical declinations outside of this range. In the statistical analysis, the declination histogram for the sample shows a pattern where the apses of the churches point consistently to the north of due east. This might signal constructions that were oriented to the rising Sun on dates close to Easter. We briefly discuss the implications these findings may have for a future study of the rest of Gran Canaria, which includes several dozens of historic churches scattered throughout the valleys, slopes, and volcanic calderas of the island.

**Keywords:** church orientation, archaeoastronomy, Christian religion.

## Introduction

The spatial orientation of historic Christian churches is one of the distinctive elements of their architecture. In Europe and in many distant places where evangelization arrived, there is a general tendency to orient the altars of these buildings in the solar range, towards those points of the horizon where the Sun rises on different days of the year, often with a clear predilection for orientations close to the geographic east (McCluskey, 2007; González-García and Belmonte, 2015). Within the same solar range, however, alignments in the opposite direction, with the altar to the west, are not unusual, although they are exceptional as they do not follow the canonical pattern (Esteban et al., 2001; Belmonte et al., 2007; Gangui and Belmonte, 2018).

The study of the layout of Christian churches has been of interest to researchers for many years. According to the texts of early Christian writers, churches were to be placed following a certain orientation, i.e., the priest had to face east during services. This was recognized by Origen, Clement of Alexandria and Tertullian, and may have been formalized during the first Council of Nicaea (325 AD). Saint Athanasius of Alexandria, also in the 4th century, expressed that the priest and the participants should turn towards the east, from where Christ, the Sun of Righteousness, would shine at the end of time (Vogel, 1962; Gangui et al., 2016). However, after the last Council of Trent (1563) there seems to be more laxity in the orientation of churches. At least, that can be inferred from the writings of Cardinal Carlo Borromeo, who participated in that council, and who later wrote his *Instructionum fabricae et supellectilis ecclesiasticae* (Borromeo, 1577).

The present paper is the continuation of a large-scale project carried out in the Iberian Peninsula and the Canary Islands. In the latter location, we have already focused on the precise orientation of the colonial churches on the islands of Lanzarote (Gangui et al., 2016), La Gomera (Di Paolo et al., 2020) and Fuerteventura (Muratore et al., 2023), as well as of those located in the city of San Cristobal de La Laguna, in the island of Tenerife (Gangui and Belmonte, 2018). Thus, it constitutes a follow up which aims to investigate whether the texts of early Christian writers and apologists, regarding the orientation of religious architecture *ad orientem*, were – or were not – respected in this limited territory, located far from the European centres of power.

## Conquest and colonization of Gran Canaria

The origins of the city of Las Palmas de Gran Canaria are in the camp organized by the captain of the Crown of Castile Juan Rejón in a palm grove next to the mouth of the Guiniguada ravine on the night of Saint John in 1478. Las Palmas can be considered the first “realengo” city (the conquest was directly sponsored by the Catholic Monarchs) founded by the Crown outside the peninsula, and an important model that was continued years later in the other Atlantic and American continent foundations. During the following decades, in various places on this island small urban centers were developed and populated and, slowly, the territory saw the emergence of estates and hamlets. In most of the villages the growing population was accompanied by the construction of small Christian chapels that reflected the new religion and social situation. In fact, the traditional way of occupying the conquered territory was through the construction of hermitages and places of worship, as a form of religious ordering of space (Delgado-Hernández and Ronquillo-Rubio, 2019). The city of Las Palmas presents a unique case to study the way in which the orientation pattern of colonial churches within compact and isolated systems (far from the metropolis) may have remained faithful to the canonical tradition, or eventually have adjusted to local needs according to weather, orography or assimilation processes.

## Historic churches of Las Palmas

The island of Gran Canaria has an abundant and rich religious architectural heritage dating from the 16th century onwards. The first Christian churches were small, simple chapels built in the different regions of the island as colonization progressed. They were, in general, buildings with a single enclosure, with a rectangular ground plan and a flat façade with a main entrance that was often the only door to access the interior.

Some chapels were located within the nascent urban centers, as is the case of the chapel of San Juan Bautista, in Las Palmas, whose construction dates back to 1672. Others were founded in more peripheral locations, such as the chapel of Nuestra Señora de la Encarnación in Tenoya (c. 1730), at some 10 km from the capital (Fig. 1).

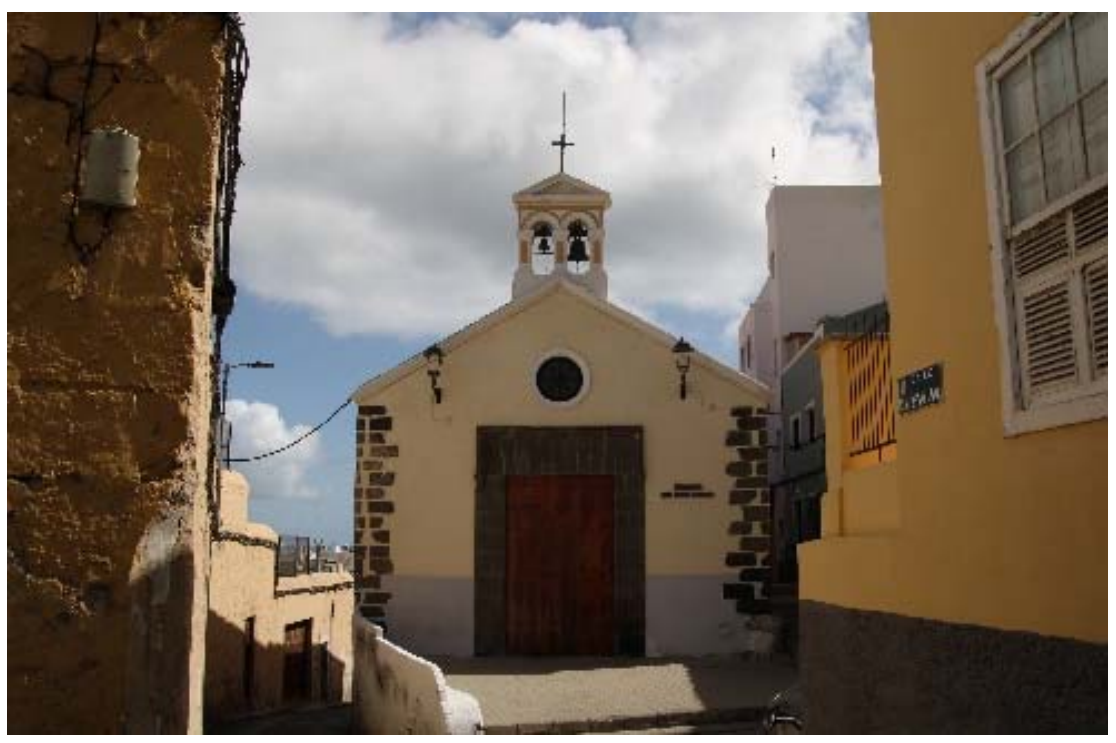
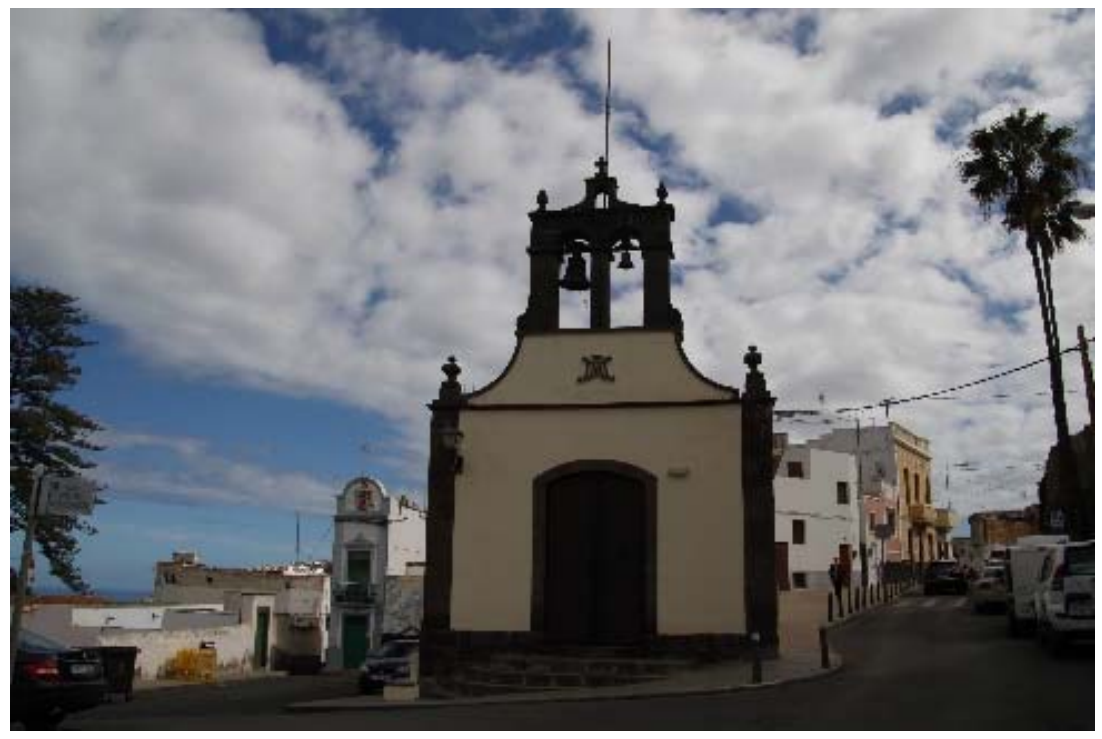

Figure 1. The chapels of San Juan Bautista, in Las Palmas (left) and Nuestra Señora de la Encarnación in Tenoya (right). Both have a rectangular ground plan and, at the front, belfries with two openings as bell towers.

Over the centuries, some of these buildings had chapels added to the chancel, sacristies on the sides or behind the altar or various ornaments on the façade. Others, due to their precarious construction that did not stand the test of time or because they were already too narrow for the number of faithful who gathered during religious ceremonies, were completely rebuilt, some of them reaching a certain monumental character. A few more representative churches located in the old part of the city of Las Palmas will be discussed below (cf. Guía del Patrimonio, 2005) together with their main characteristics and orientations.

**Data sample, methods, and some results**

We undertake a systematic study of the orientation of the Christian churches of the city of Las Palmas and its surroundings. Our main interest now is to conduct a statistical analysis of the sample, as this can provide us with unique archaeoastronomical data comprising a compact set of old churches and chapels where we can search for Catholic religious traditions and even for pre-European ones (Belmonte, 2015), including astronomical, or a mix of both. As with previous works, this may offer us a broader understanding of a key aspect of Canarian culture.

Figure 2 shows the geographical location of all the churches and chapels measured (marked with numbers according to Table 1). We obtained our measurements using a tandem instrument Suunto 360PC/360R, which incorporates a clinometer and a compass with a precision of 0.5°, and by analyzing the surroundings of each of the buildings. We then corrected the azimuth data according to the local magnetic declination (Natural Resources Canada, 2023). As further corroboration of our measurements, especially for the biggest churches, the orientations we obtained *in situ* were later verified with satellite photo-images. Our data is the result of several on-site measurements with a single instrument, taking the axes of the churches, from the back of the buildings towards the altars, as our main guide. In many cases, especially for the small constructions, we could also verify that the lateral walls were parallel to their axes.

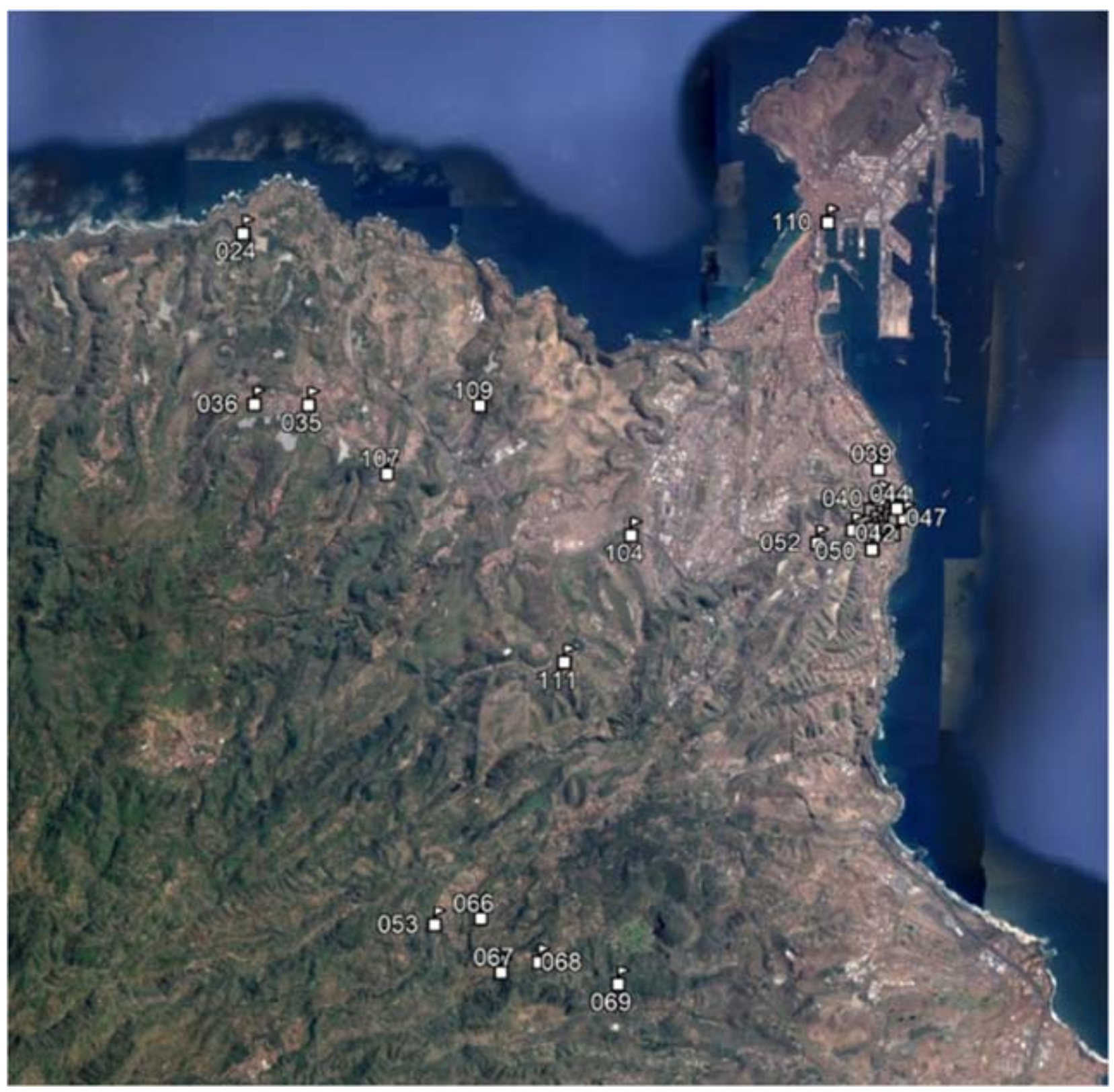

Figure 2. Map of the region surrounding the city of Las Palmas de Gran Canaria with the geographical location of the full sample of churches measured. Image on a map of Google Earth.

Table 1 lists the identification of the churches, along with their geographical location and archaeoastronomical data: the measured azimuth and the angular height of the point of the horizon towards which the altar of the church is facing, as well as the derived computed declination corresponding to the central point of the solar disc (Ruggles, 2015a; 2015b). The measured height of the horizon was appropriately corrected for atmospheric refraction (Schaefer, 1993) and when the horizon was blocked, we employed the digital elevation model (DEM) based on the Shuttle Radar Topographic Mission (SRTM) available at (Kosowsky, 2023), which gives angular heights within a 0.5° approximation. The uncertainty associated with each value of declination was obtained by error propagation (Ruggles, 2015a), which gave a mean value of 0.7°.

Table 1: Orientations for the chapels and churches of Las Palmas, ordered by increasing azimuth. For each building, we show the location, identification, the geographical latitude and longitude (L and l), the astronomical azimuth (a) taken along the axis of the building towards the apse, the horizon angular height (h) in that direction (including the correction due to atmospheric refraction; B means the horizon was blocked and we used DEM data), both rounded to 1/2º approximation and expressed in decimal degrees, and the corresponding resultant declination (δ). The last column indicates the Patron saint date of each church and the Orientation (Gregorian date unless Julian is specified), which is computed by considering the approximate year of the construction of each church (signaled in the second column) and then

estimating the dates when the declination of the Sun is the one indicated. The numbers in the first column correspond to those signaling the geographical locations of the churches in the map of Figure 2.

| Location | Name (date) | L (º, N) | l (º, W) | a (º) | h (º) | δ (º) | Patron saint date / Orientation |
|---|---|---|---|---|---|---|---|
| 045 Las Palmas | Santa Ana (Catedral, c. 1500) | 28.100677 | 15.415221 | 62.3 | B –0.2 | 24.1 | 26 Jul / 10-15 Jun (Julian) |
| 053 Santa Brígida | Santa Brígida (1756) | 28.033501 | 15.499831 | 68.9 | B +2.0 | 19.5 | 1 Feb / 15 May-28 Jul |
| 109 Tenoya | Ntra. Sra. de la Encarnación (c. 1730) | 28.118734 | 15.491437 | 69.6 | +1.0 | 18.4 | 25 Mar / 11 May-3 Aug |
| 040 Las Palmas | San Francisco de Asís (1518) | 28.103374 | 15.417207 | 71.0 | B –0.2 | 16.6 | 4 Oct / 25 Apr-31 Jul (Julian) |
| 111 San Lorenzo | San Lorenzo (1681) | 28.076517 | 15.475728 | 71.6 | +5.7 | 18.9 | 10 Aug / 12 May-31 Jul |
| 107 Santidad | Ntra. Sra. del Carmen | 28.107526 | 15.508675 | 74.7 | –0.1 | 13.4 | 16 Jul / 24 Apr-20 Aug |
| 039 Las Palmas | San Pedro González Telmo (XVII c.) | 28.108248 | 15.417280 | 75.8 | B –0.2 | 12.4 | 14 Apr / 21 Apr-22 Aug |
| 067 Santa Brígida | Ntra. Sra. de la Inmaculada Concepción (1733) | 28.025676 | 15.487527 | 76.5 | –0.6 | 11.6 | 8 Dec / 19 Apr-25 Aug |
| 048 Las Palmas | Cristo del Buen Fin (Espíritu Santo, 1615) | 28.100021 | 15.416146 | 80.7 | –0.2 | 8.1 | ?? / 9 Apr-4 Sep |
| 044 Las Palmas | San Antonio Abad (1757) | 28.101768 | 15.413878 | 83.4 | B –0.2 | 5.7 | 17 Jan / 3 Apr-10 Sep |
| 069 Las Goteras | Ntra. Sra. del Carmen (c. 1737) | 28.023706 | 15.465747 | 84.4 | B –0.5 | 4.7 | 16 Jul / 31 Mar-12 Sep |
| 042 Las Palmas | San Roque (1939) | 28.098217 | 15.422134 | 85.7 | +0.3 | 3.9 | 16 Aug / 30 Mar-15 Sep |
| 047 Las Palmas | San Agustín (1786) | 28.100163 | 15.412599 | 86.5 | B –0.2 | 3.0 | 28 Aug / 27 Mar-17 Sep |
| 043 Las Palmas | San Juan Bautista (1672) | 28.098560 | 15.418562 | 92.0 | –0.2 | -1.9 | 24 Jun / 14 Mar-29 Sep |
| 049 Las Palmas | Santo Domingo de Guzmán (1841) | 28.098665 | 15.415967 | 93.2 | B –0.2 | -2.9 | 8 Aug / 12 Mar-2 Oct |
| 110 Las Palmas | Ntra. Sra. de la Luz (c. 1796) | 28.148709 | 15.426692 | 98.3 | B +0.2 | -7.2 | 2nd Saturday Oct / 29 Feb-12 Oct |
| 035 Arucas | San Juan Bautista (1909) | 28.118829 | 15.523258 | 106.4 | –0.3 | -14.6 | 24 Jun / 9 Feb-4 Nov |
| 050 Las Palmas | San José (c. 1788) | 28.095002 | 15.418526 | 174.1 | +0.3 | -61.1 | 19 Mar / ---- |
| 046 Las Palmas | San Francisco de Borja (1724) | 28.099971 | 15.414429 | 176.3 | B +0.5 | -61.2 | 3 Oct / ---- |
| 052 Las Palmas | Ntra. Sra. del Rosario de Fátima | 28.096097 | 15.428711 | 215.0 | +4.4 | -43.2 | 13 May / ---- |
| 041 Las Palmas | San Nicolás de Bari (1697) | 28.101591 | 15.418821 | 228.5 | +4.2 | -33.3 | 6 Dec / ---- |
| 068 La Atalaya de Sta. Brígida | San Pedro Apóstol (c. 1737) | 28.027316 | 15.480484 | 246.1 | +6.3 | -17.7 | 29 Jun / 29 Jan-14 Nov |
| 024 Bañaderos | San Pedro Apóstol (1877) | 28.146938 | 15.535424 | 263.0 | +8.8 | -2.0 | 29 Jun / 14 Mar-30 Sep |
| 036 Arucas | Cristo de la Salud (XVIII c.) | 28.118967 | 15.533358 | 270.5 | +5.3 | 2.9 | 14/15 Sep / 27 Mar-17 Sep |
| 051 Las Palmas | Ntra. Sra. de los Reyes (c. 1610) | 28.097494 | 15.414549 | 286.3 | +3.0 | 15.8 | 15 Aug / 2 May-12 Aug |
| 066 San José de las Vegas | San José (1711) | 28.034501 | 15.491293 | 296.0 | +3.6 | 24.6 | 19 Mar / ~Solst. Jun. |
| 104 La Mayordomía | San Antonio Abad (XVIII c.) | 28.097376 | 15.463332 | 317.6 | +2.0 | 41.9 | 17 Jan / ---- |

For our analysis, we computed a declination curvigram that provides information on the probability of finding a particular value of the astronomical declination within our sample, following the methodology of several preceding works (e.g., González-García et al., 2021). We first calculated the probability density function of the observed sample. For this, we used an appropriate smoothing using a Gaussian kernel function, considering a bandwidth equal to twice the average uncertainty of the declination measurements.

To find out if a concentration of declination values is significant, we compared the probability distribution of the observed sample against a set of probability distributions with declinations calculated from random azimuth values distributed between 0° and 360°, and a mean latitude of 28°05', which is close to the center of the historic Vegueta quartier.

We used 230 random distributions, each with the same amount of data as the observed sample. We then scaled our observed distribution with respect to those random ones to check whether the measures deviated significantly or were confounded by the latter. The significance level was estimated by subtracting from the observed probability distribution the average probability of the random distributions, and then dividing the result (point by point) by the corresponding standard deviation.

Due to the similarity of this calculation with those current in some areas of physics, the result can be called a signal-to-noise ratio (SNR), where the scale is given by the standard deviation (σ) and, therefore, any peak that rises above 3σ is statistically significant – i.e., can be considered incompatible with a random sample at a 99% confident level – according to our analytical approach, as we show in Fig. 3. This provides a precise view of the concentration or the probability density of certain orientation patterns that might be a matter of interest.

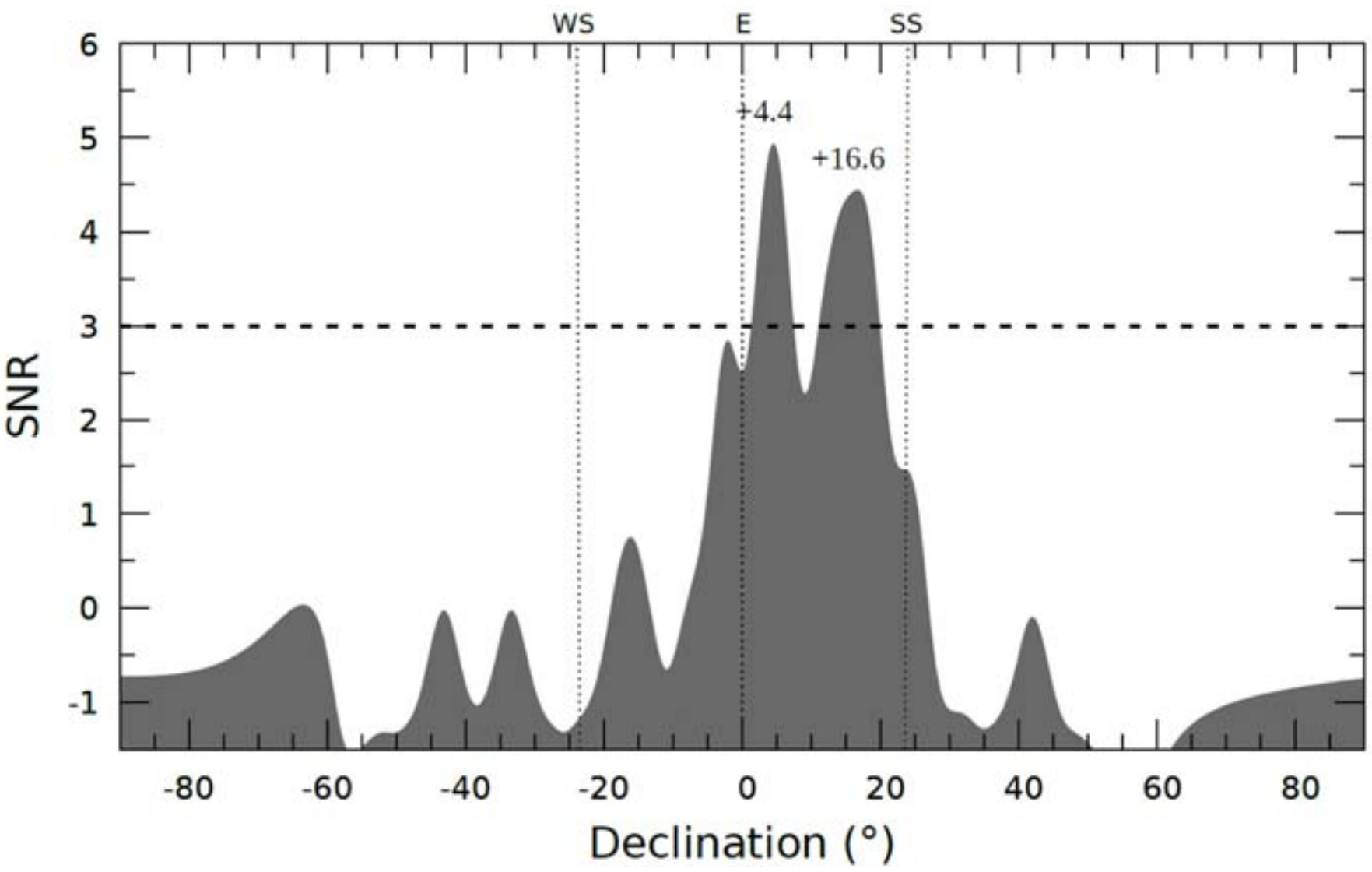

Figure 3. Normalised declination curvigram (or SNR) of the churches analysed. The astronomical equinox (E) and the winter (WS) and summer (SS) solstices are marked with vertical dotted lines. The horizontal dashed line represents the 3σ level and is a direct outcome of our normalisation process. Peaks rising above this line can be considered potentially significant.

As can be seen from the plot, the normalized curvigram shows a preference for orientations within the solar range. The principal maximum in Figure 3 appears at a declination of c. +4.4º, but the peak is broad enough to also include the equinox at 0º. A second broad peak rising above the 3σ level is located at c. +16.6º.

**Discussion and future directions**

Table 1 of the previous section shows that most of the churches' orientations of the region explored fit within the solar arc. Furthermore, the declination curvigram of Fig. 3 helps us appreciate the significance of the largest concentrations in the estimated declination values. Therefore, the great majority of the constructions' altars (22 out of the 27 we measured) face the direction of sunrise (17) or sunset (5) at some time in the year. This orientation conforms with Christian tradition (McCluskey, 2015).

However, the actual pattern in the declination plot is not easy to explain. The usual theory of orientations toward the sunrise on the church patron's feast day was discarded by individual evaluation, as can be seen from Table 1, in which no patron saint's day fits with the corresponding astronomical orientation. Moreover, we find no correlation between the global orientation pattern and the orography of the studied region surrounding the city of Las Palmas. One could have expected the natural Guiniguada ravine to direct roughly the azimuthal disposition of old churches close to it, but no hint of this is revealed in our data.

Interesting singular examples are San Antonio Abad (dated 1757), located in the old Vegueta quartier of the capital city, San Pedro González Telmo (XVII c.), built originally close to the sea, and Santa Ana (c. 1500), the cathedral of Las Palmas.

San Antonio, in particular, warns us to be careful with dates and pay close attention to possible church reconstructions. As tradition says, Christoph Columbus prayed in this church, possibly an older construction than the current one which is from the 18th century. The value of declination of 5.7º for this church is potentially Julian equinoctial (Solar declination on the 21 March at the time of Columbus voyage is 4.5º, not too different from the measured 5.7º). And we should not forget the relevance of the "equinox" for the aboriginal population (Esteban et al., 1996; 1997). In summary, this church in Vegueta, if it was rebuilt on top of the old, original, church that Columbus visited, could be oriented towards the ecclesiastical equinox.

San Pedro González Telmo has an azimuth close to 76º and fits within the solar range. However, for the time being one cannot rule out that this orientation responded more to the fact that it was perpendicular to the original coast (in our days, the coastal lands were filled and used for a bus station and a highway) than that it obeyed a particular solar orientation. Something similar applies for San Agustín (1786), originally located close to the shore and oriented just a few degrees north of due east.

Lastly, let us consider Santa Ana, the city's Cathedral, which has a clear solstitial orientation. This result is suggestive, since it marks the summer solstice, "quando el Sol entra en Cancro a veinte y uno de junio en Adelante" ("when the Sun enters Cancer on the twenty-first of June onwards"), according to the historian Marín de Cubas (1694), the most important date in the aboriginal calendar. Let's not forget that the Moon following this date was the one for festivals for the aboriginal population, and there are many different solstitial markers on the island of Gran Canaria, for example, the one located on the Cuatro Puertas mountain, in the municipality of Telde, an artificial cave with four accesses carved into the volcanic tuff. This orientation towards sunrise on the summer solstice, although subtle, could perhaps crystallize the Christianization of a rite prior to the arrival of the conquerors. Although this is only an indication – which must be studied in depth in the future for the rest of the island – we can recall that, around the same time, a very similar situation would occur in another part of the Canaries, in San Cristóbal de La Laguna, this time with the western solstitial orientation of two very relevant churches, Nuestra Señora de la Concepción and San Agustín, both from the beginning of the 16th century (Gangui and Belmonte, 2018).

Let us now consider the main peak in the curvigram of Figure 3, at a declination of c. +4.4º, slightly to the north of due east. This peak might suggest that a not negligible group of churches in the region we studied were not oriented to the ecclesiastical equinox on 21 March, as canonical texts indicate, but to the sunrise on Easter Sunday at the approximate year of the construction of the building. This is indeed one of the most important feast days of Christianity and it is not the first time that a similar situation occurs (e.g., Urrutia-Aparicio et al., 2021; Muratore et al., 2023). To test this hypothesis, we consider all churches with declination values within some five degrees on both sides of the +4.4º peak and check the possible Easter dates matching their declination (see Table 2).

Table 2: List of chapels and churches with declination within +4.4±5º. The table reproduces data of Table 1, including a new last column. This column indicates the dates of Easter Sundays on years that are close to the most likely years of construction when the declination of the Sun is approximately the corresponding declination for each church.

| **Location** | **Name (date)** | **δ (º)** | **Closest Easter date** |
|---|---|---|---|
| 048 Las Palmas | Cristo del Buen Fin (Espíritu Santo, 1615) | 8.1±0.7 | 11 Apr 1610 (**δ**=8.8)<br>7 Apr 1613 (**δ**=7.4) |
| 044 Las Palmas | San Antonio Abad (1757) | 5.7±0.7 | 2 Apr 1752 (**δ**=5.6)<br>3 Apr 1763 (**δ**=5.7) |
| 069 Las Goteras | Ntra. Sra. del Carmen (c. 1737) | 4.7±0.7 | 1 Apr 1736 (**δ**=5.2)<br>2 Apr 1741 (**δ**=5.5) |
| 042 Las Palmas | San Roque (1939) | 3.9±0.7 | 28 Mar 1937 (**δ**=3.3) |
| 047 Las Palmas | San Agustín (1786) | 3.0±0.7 | 26 Mar 1780 (**δ**=2.9)<br>27 Mar 1785 (**δ**=3.3) |
| 036 Arucas | Cristo de la Salud (XVIII c.) | 2.9±0.7 | 25 Mar 1742 (**δ**=2.2)<br>26 Mar 1758 (**δ**=2.7) |

We see from Table 2 that given the value of a church's declination it is not difficult to find a Sun's declination close to it but occurring during Easter Sunday in a small range of years around the construction date. However, we should also bear in mind that, as we already mentioned, this is a broad peak which also includes the equinox declination at 0º (and there is even another secondary peak very close to the 3σ level at declination c -2º, to confuse things even more). So, as it happened with the declination measured for San Antonio Abad, one cannot exclude a priori the possibility that this accumulation of declinations in the plot is just a signature of equinoctial orientations.

Regarding the peak at a declination c. +16.6º in the curvigram of Figure 3, we see from Table 1 that roughly half of the churches contributing to it are located relatively far from the capital city (although San Francisco de Asís, in Triana, belongs to Las Palmas). At present we do not have a clear explanation for this peak (neither proximity, related cult, orography, local calendar feasts or an obvious astronomical reason apply), which might be a rare but interesting case of a spurious statistically significant peak. However, it is likely that when we complete our study of the island, thus completing the sample, a new hint may arise.

To conclude, in the future these studies should be completed with the measurements and analysis of the orientation of the many heritage churches that are distributed all over the island and, in particular, in the UNESCO Biosphere Reserve, a protected area which covers 46% of the territory in its western half. One might expect – especially in areas far from the first Castilian settlements – to find traces of orientations more typical of the aboriginal inhabitants of the island, who had customs and rites very different from those of the new occupants of the territory.

**Acknowledgements**

A.G. is grateful for the support of the Cabildo de Gran Canaria. He thanks Carlos Santana for his constant help in preparing the visit to the island and acknowledges further support from CONICET (PIP 11220210100111CO) and the University of Buenos Aires (grant 20020190100160BA).